\documentclass[aps,a4paper,11pt,preprint,notitlepage]{revtex4} 
 
\usepackage{color}
\usepackage{graphicx}
\usepackage{braket}
\usepackage{amsmath}
\usepackage{enumitem}
\usepackage{hyperref}
\usepackage{ulem}

\definecolor{newgreen}{rgb}{0,.5,0}

\newcommand{\be}{\begin{equation}}
\newcommand{\ee}{\end{equation}}
\newcommand{\ba}{\begin{eqnarray}}
\newcommand{\ea}{\end{eqnarray}}

\begin{document} 
 
\title{A Quantum-Classical Model of Brain Dynamics} 

\author{Alessandro Sergi} 
\affiliation{Dipartimento di Scienze Matematiche e Informatiche,
Scienze Fisiche e Scienze della Terra,
Universit\`a~degli~Studi~di~Messina,s
viale F. Stagno d'Alcontres 31, 98166 Messina, Italy;
E-mail: asergi@unime.it}
\affiliation{Institute of Systems Science, Durban University of Technology,
P.O. Box 1334, Durban 4000, South Africa}

\author{Antonino Messina}
\affiliation{Dipartimento di Matematica ed Informatica, Universit\`a
degli Studi di Palermo, Via Archirafi 34,
90123 Palermo, Italy; E-mail: antonino.messina1949@gmail.com}

\author{Carmelo M. Vicario}
\affiliation{Dipartimento di Scienze Cognitive, Psicologiche,
Pedagogiche e degli Studi Culturali,
Universit\`a~degli~Studi~di~Messina,
Via Concezione 6,
98121 Messina, Italy; E-mail: carmelomario.vicario@unime.it}

\author{Gabriella Martino}
\affiliation{Dipartimento di Medicina e Clinica Sperimentale,
Universit\`a degli Studi di Messina,
A.O.U. ``G. Martino'', Via Consolare Valeria,
98125 Messina, Italy; E-mail: gabriella.martino@unime.it}

\begin{abstract}
The study of the human psyche has elucidated a 
bipartite structure of logic reflecting
the quantum--classical nature of the world. Accordingly,
we posited an approach toward studying the brain by means of the quantum--classical dynamics 
of a mixed Weyl symbol. The mixed Weyl symbol
can be used to describe brain processes
at the microscopic level and, when averaged over an appropriate ensemble, 
can provide a link to the results of measurements made at the meso 
and macro scale.
Within this approach, quantum variables (such as, for example,
nuclear and electron spins, 
dipole momenta of particles or molecules, tunneling degrees of
freedom, and so on) can be represented by spinors,
whereas the electromagnetic fields and phonon modes
can be treated either classically or semi-classically
in phase space by also 
considering quantum zero-point fluctuations.  
Quantum zero-point effects can be incorporated into numerical simulations
by controlling the temperature of each field
mode via coupling to a dedicated Nos\'e-Hoover chain thermostat.
The temperature of each thermostat was chosen in order to
reproduce quantum statistics in the canonical ensemble.
In this first paper, we introduce a general quantum--classical
Hamiltonian model that can be
tailored to study physical processes
at the interface between the quantum and the classical world in the brain.
While the approach is discussed in detail, numerical calculations
are not reported in the present paper, 
but they are planned for future work.
Our theory of brain dynamics subsumes some compatible aspects
of three well-known quantum approaches to brain dynamics,
namely the electromagnetic
field theory approach, the orchestrated objective reduction theory, 
and the dissipative quantum model of the brain.
All three models are reviewed.
\end{abstract}

\maketitle

\section{Introduction}
\label{sec:intro}

The human brain is perhaps the most complicated known condensed matter system. It contains approximately $10^2$ billions of neurons
and at least as many glia cells~\cite{vonbartheld}.
The brain is composed of 77 to 78\% water, 10 to 12\% lipids,
8\% proteins,  2\% soluble organic substances, 
and 1\% carbohydrates and inorganic salts~\cite{McIlwain}.
It is also extremely fascinating that higher brain functions
precisely define what it means to be human.
Brain states and their dynamics have so far eluded
physical understanding based on molecular models. 
This means that 
one cannot describe brain dynamics by brute force,
i.e.,~starting from the behavior of all atoms 
and deriving macroscopic time evolution.
The problem does not only reside in the sheer
number of microscopic constituents of the brain.
Some of the most complex brain functions are delocalized
over long distances and require synchronization processes
that do not seem easy to explain only by means of the
classical mechanics of atoms and molecules.
In particular, the~wholeness of perception
requires integrating the activity of an enormous number of brain cells.
Ultimately, we would like to develop a theory
of brain processes where mesoscopic models can be constructed
from atomistic dynamics by means of controlled approximations.
With respect to this, quantum models
~\cite{mcfadden,mcfadden2,mcfadden3,pockett,pockett2,pockett3,
liboff,liboff2,frohlich,ph,ph-2,ph-3,penrose,penrose2,penrose3,
penrose4,penrose5,microtubulines_channels,microtubulines_channels-2,
fisher,fisher2,fisher4,
kerskens,qc-bneuro-comp,vitiello1995,pessa-vitiello,alfinito-vitiello,
freeman-vitiello,freeman-vitiello2,freeman-vitiello3,vitiello-fractals,
vitiello-cortex,sabbadini-vitiello}
may hold the key to a possible microscopic understanding of
some brain~functions.

We have found that some psychological theories are based
on a bi-partite logic~\cite{science_and_sanity,drive_yourself_sane,collected,
stealing_general_semantics,pauli,fuchs,lindorff,synchronicity,jung-pauli,
blanco,blanco2,rayner,lombardi} that is very similar to the logic of
quantum--classical mechanics.
We do wish to make clear at the very beginning
that, in this paper,
the words ``psyche'', ``psychological'',
and the like are not used to address any metaphysical level of `reality'.
While we acknowledge that such concepts lack, at
the moment, both quantitative definitions and complete explanations in terms
of biomolecular processes, it must be emphasized that neuroscience
~\cite{neuroscience} is continuously advancing toward the
inclusion of psychological
phenomena within the boundaries of quantitative science.
Thus, once our use of these words is understood, it becomes easier to
accept the idea that psychological theories and clinical psychology
could feed the synergistic growing of translational neuroscience
~\cite{trans1,trans2,trans3,trans4,trans5,trans6,trans7},
quantum models of decision-making~\cite{tanaka,khrennikov,khrennikov2,busemeyer-book,khrennikov3,
bond,basieva,busemeyer,vandernoort},
and quantum information biology~\cite{qib}.

In recent years, three main quantum models of the
brain have been introduced in the literature. These are
the electromagnetic field (EMF) approach
~\cite{mcfadden,mcfadden2,mcfadden3,pockett,pockett2,pockett3,
liboff,liboff2,frohlich},
the orchestrated objective reduction (Orch OR) theory~\cite{ph,ph-2,ph-3,penrose,penrose2,penrose3,penrose4,penrose5,
microtubulines_channels,microtubulines_channels-2,fisher,fisher2,
fisher4,kerskens,qc-bneuro-comp},
and the dissipative quantum model of brain (DQMB) 
\cite{vitiello1995,pessa-vitiello,alfinito-vitiello,freeman-vitiello,
freeman-vitiello2,freeman-vitiello3,vitiello-fractals,vitiello-cortex,
sabbadini-vitiello}.
Even if there are several key differences between the EMF, the~Orch OR
approaches, and~DQMB, these three theories study the brain
from the perspective
of condensed matter physics and matter--EMF interactions.
While DQMB~\cite{vitiello1995,pessa-vitiello,alfinito-vitiello,freeman-vitiello,
freeman-vitiello2,freeman-vitiello3,vitiello-fractals,
vitiello-cortex,sabbadini-vitiello} is mainly concerned with the 
explanation of memory storage and retrieval, long-range correlations
between brain clusters of cells and brain correlates of perception,
both the EMF~\cite{mcfadden,mcfadden2,mcfadden3,pockett,pockett2,pockett3,
liboff,liboff2,frohlich} and Orch OR
~\cite{ph,ph-2,ph-3,penrose,penrose2,penrose3,penrose4,penrose5,
microtubulines_channels,microtubulines_channels-2,fisher,fisher2,
fisher4,kerskens,qc-bneuro-comp}
models were originally introduced for explaining consciousness.
With regard to this, we want to state very clearly that
our theory does not aim in any way to explain consciousness.
Instead, we stress that our target is only to study brain dynamics
in terms of physical
processes. In~practice, we only consider EMF
~\cite{mcfadden,mcfadden2,mcfadden3,pockett,pockett2,pockett3,
liboff,liboff2,frohlich} and Orch OR~\cite{ph,ph-2,ph-3,penrose,penrose2,penrose3,penrose4,penrose5,
microtubulines_channels,microtubulines_channels-2,fisher,fisher2,
fisher4,kerskens,qc-bneuro-comp} 
insofar that they can be used
as microscopic theories of physical processes in the~brain.

Motivated by the germinal considerations in Refs.
~\cite{mcfadden,mcfadden2,mcfadden3,pockett,pockett2,pockett3,
liboff,liboff2,frohlich,ph,ph-2,ph-3,penrose,penrose2,penrose3,
penrose4,penrose5,microtubulines_channels,microtubulines_channels-2,
fisher,fisher2,fisher4,
kerskens,qc-bneuro-comp,vitiello1995,pessa-vitiello,alfinito-vitiello,
freeman-vitiello,freeman-vitiello2,freeman-vitiello3,vitiello-fractals,
vitiello-cortex,sabbadini-vitiello} and by 
the idea of a bipartite structure of logic
~\cite{science_and_sanity,drive_yourself_sane,collected,
stealing_general_semantics,pauli,fuchs,
lindorff,synchronicity,jung-pauli,blanco,blanco2,rayner,lombardi},
in this paper, we introduce an explicit quantum--classical model of brain dynamics.
Such a model is based on the hybrid quantum--classical (QC) formalism
of Refs.~\cite{silin,rukhazade,balescu,zhang-balescu,balescu-zhang,aleksandrov,
gerasimenko,boucher,petrina,prezhdo,kapracicco,nielsen,as4,as6,as14,
osborn,martens,donoso,as3,as13,as7,ak}.
In many QC theories, the~nature of the interaction between the
classical and quantum subsystems is somewhat unclear and the quantum
variables are not treated on the same footing as
the classical DOF. The~formulation of Refs.
~\cite{silin,rukhazade,balescu,zhang-balescu,balescu-zhang,aleksandrov,
gerasimenko,boucher,petrina,prezhdo,kapracicco,nielsen,as4,as6,as14,
osborn,martens,donoso,as3,as13,as7,ak}
is based on mixed Weyl symbols and is conceptually free from these drawbacks. 
In fact, such an approach is founded upon
a statistical operator depending parametrically on phase space points.
This implies that the dynamics must be considered at each phase space
point without the possibility of separating quantum dynamics from the
classical-like dynamics of the phase space.
QC spin--boson models~\cite{leggett,bakermeier,plenio,finney},
and their non-linear extension~\cite{ak}, are appropriate for describing
a finite number of quantum variables coupled to a classical DOF.
Non-Hamiltonian deterministic thermostats~\cite{as4,as6,as14,nhc,b1,b2}
can be used to formulate the dissipative dynamics of mixed Weyl symbols
under constant temperature conditions.
The QC formalism simplifies numerical calculations
of averages and response functions. In~turn, response functions can be compared
to electromagnetic signals that could be provided by
macroscopic experiments on the brain
~\cite{riddle,riddle2,abubaker,croce,caruana,nitsche-paulus,stagg-nitsche,
papazova,yavari,frohlich,koch,martinez-banaclocha,pinotsis,
vicario2020a,ney2021,markovic2021,vicario2020b,nunez}.

We are interested in studying those brain processes that
can be described in terms of a few quantum variables
embedded in a classical environment.
Small numbers of quantum particles are naturally found in
small biological structures
~\cite{microtubulines_channels,microtubulines_channels-2},
and from such a scale until that of atomic nuclei
~\cite{fisher,fisher2,fisher4}.
Even a small number of quantum variables can have a significant
effect on the dynamics of large classical systems by means of four mechanisms.
One is given by non-adiabatic transitions between energy states
~\cite{osborn,martens,donoso,as3,as13,as7,ak}.
The second one is caused by the stochastic collapse of the 
wave function~\cite{stapp,vonneumann}. The~third one is generated 
by the motion of quantum sources of the electromagnetic fields
in the brain. The~fourth one is the famous `order from order'
mechanism elaborated on by Schr\"odinger~\cite{what_is_life},
which led to the discovery of DNA~\cite{dna,dna2}.
All of these mechanisms are in agreement with
Pascual Jordan's idea~\cite{beyler,beyler2,mcfadden-alkhalili}
about the necessary role of the amplification of quantum
processes in order to steer classical dynamics
in biological environments. Hence, one can consider that
single quantum particles, such as electrons and protons, retain 
quantum \mbox{properties~\cite{osborn,martens,donoso,as3,as13,as7,ak}}
at every temperature~\cite{mcfadden-book}.
For such a reason, our QC approach can cope `almost by design'
with the controversial issue of decoherence
~\cite{joos,zurek-2003,tegmark,microtubulines_channels} in warm
and wet environments, such as those found in biological systems.
It has also been proposed that
some kind of quantum computation~\cite{nielsen-chuang,jaeger}
might take place in the brain~\cite{microtubulines_channels,fisher,fisher2,kerskens,qc-bneuro-comp}.
There are a few proposals regarding quantum computational schemes
performed by means of mixed states
~\cite{no-ent-comp,no-ent-comp2,no-ent-comp3,no-ent-comp4};
however, the~mainstream concept of quantum computation
requires entangled states~\cite{nielsen-chuang,jaeger}.
Entanglement is fragile and, for~quantum computation to
be realized at biological conditions,
decoherence~\cite{joos,zurek-2003} should not destroy
the phase coherence of quantum states at a high temperature
~\cite{tegmark}. Such a possibility is still a matter of debate
~\cite{microtubulines_channels}.
For such a reason, 
in this paper, we do not take a quantum informational perspective
~\cite{nielsen-chuang,jaeger}.

The paper is structured as follows.
We present the historic evolution of logic's bipartite structure
by discussing general semantics (GS) in Section~\ref{sec:gs},
synchronicity in Section~\ref{sec:jung},
and Blanco's bi-logic in Section~\ref{sec:blanco}.
In Section~\ref{sec:jordan}, we discuss a set of ideas
in favor of quantum mechanical effects in the brain.
We review the EMF approach in Section~\ref{sec:emf}, 
Orch OR in Section~\ref{sec:penrose},
and DQMB in Section~\ref{sec:umezawa}.
Our QC approach is presented in Section~\ref{sec:qcdmb}.
Finally, our conclusions are given in Section~\ref{sec:end}.

\section{The Bipartite Structure of Psychology
as the Root for Quantum--Classical Models of the~Brain}
\label{sec:psyche}

The knowledge that some theories of the human psyche
suggest that logic has a bi-partite structure
~\cite{science_and_sanity,drive_yourself_sane,collected,
stealing_general_semantics,pauli,fuchs,lindorff,synchronicity,jung-pauli,blanco,
blanco2,rayner,lombardi}, paralleling that of the quantum--classical world,
is, for us, one of the inspiring motivations to take the first steps
toward the elaboration of a quantum--classical model of brain dynamics.
As is clarified in the following,
bi-logic comprises Aristotelian and non-Aristotelian logics.
Discussions on the bi-partite structure of logic 
can also be of interest to quantum models of decision making~\cite{tanaka,khrennikov,khrennikov2,busemeyer-book,khrennikov3, 
bond,basieva,busemeyer,vandernoort},
quantum information biology~\cite{qib},
and translational neuroscience
~\cite{trans1,trans2,trans3,trans4,trans5,trans6,trans7}.

In the field of clinical psychology, Korzybski 
was one of the pioneers making use of non-Aristotelian logic
~\cite{science_and_sanity,drive_yourself_sane,collected,
stealing_general_semantics} for therapeutical applications.
A more abstract and somewhat implicit approach to
such a bi-partite structure of logic can be found in the work 
of Jung and Pauli~\cite{pauli,fuchs,lindorff,synchronicity,jung-pauli}.
Instead, the~most complete formulation of bi-logic
and its application to clinical psychology
(until now) is found in the work of Blanco
~\cite{blanco,blanco2,rayner,lombardi},
where it is called bi-logic.
Since, in QM, the law of the excluded middle is not valid,
so {that, as} in the famous Schr\"odinger's example
~\cite{cat}, a~cat can be both alive and dead~\cite{nielsen-chuang,jaeger},
at a fundamental level,
quantum logic is non-Aristotelian, and we propose to identify it
with Blanco's bi-logic.  If~the quantum world does not follow Aristotelian logic,
given that QM cannot indeed be separated
from the classical world, because~of the processes of
`measurement' and the stochastic collapse of the wave function
~\cite{stapp,vonneumann},
then physical processes must have a hybrid structure,
where both Aristotelian and non-Aristotelian logic
must be employed. Thus, we sustain that quantum physical processes
can be described by hybrid quantum--classical models.
Since bi-logic~\cite{blanco,blanco2,rayner,lombardi}
can be likened to a quantum--classical worldview
and there are already models supporting the quantum--classical
nature of the brain
~\cite{mcfadden,mcfadden2,mcfadden3,pockett,pockett2,pockett3,
liboff,liboff2,frohlich,ph,ph-2,ph-3,penrose,penrose2,penrose3,
penrose4,penrose5,microtubulines_channels,microtubulines_channels-2,
fisher,fisher2,fisher4,
kerskens,qc-bneuro-comp,vitiello1995,pessa-vitiello,alfinito-vitiello,
freeman-vitiello,freeman-vitiello2,freeman-vitiello3,vitiello-fractals,
vitiello-cortex,sabbadini-vitiello},
the idea of developing models by means of
an explicit quantum--classical theory naturally arises.
Our parallelism between bi-logic~\cite{blanco,blanco2,rayner,lombardi}
and quantum--classical phenomena in the brain
can also be considered as the motivation for extending the current quantum-like
models of cognition
~\cite{tanaka,khrennikov,khrennikov2,busemeyer-book,khrennikov3,
bond,basieva,busemeyer,vandernoort}
to take into account quantum--classical processes.
In the remaining part of this section, we discuss the historical
development of non-Aristotelian logic in psychology and clinical
psychology.

\subsection{General~Semantics}
\label{sec:gs}

Roughly speaking, GS is a specific instance of
clinical psychology with the specific goal
of improving mental health and adaptation to the world~\cite{science_and_sanity,drive_yourself_sane}.
One key aspect of this approach is the claim that
non-Aristotelian logic conforms more to reality.
Once non-Aristotelian logic is accepted as the correct way of thinking,
our language must be adjusted accordingly.
GS was created with the logical structure of QM
as a template~\cite{science_and_sanity}.
An interesting connection of GS to quantum models
of decision making, which is yet to be fully explored,
may be founded on the free energy principle~\cite{friston,sanchez-canizares}.
However, a~first application of this principle
to quantum decision making can be found in Ref.~\cite{tanaka}.

Although it constitutes
the historical roots of many systems of clinical psychology,
GS is rarely acknowledged~\cite{stealing_general_semantics}.
The premises of GS are ``A map is not the territory'',
``A map does not represent all of a territory'', and~``A map is self-reflexive'', meaning that an 'ideal' map would
include a map of the map, etc., indefinitely'' \cite{collected}.
These assumptions can be translated to daily life in order
to improve the mental sanity of human beings~\cite{drive_yourself_sane}.
In this case, GS premises become ``A word is not what it represents'',
``A word does not represent all of the {\rm facts}'', 
and ``Language is self-reflexive'' in the sense that, in language, we
can speak about language.
Alas, human being reactions to verbal communication are largely based
on unconscious beliefs, violating the first two assumptions and 
disregarding the third.
Mathematics and GS are the only languages that rigorously
take into account the above non-Aristotelian premises at all times.
For such a reason, Korbizski strongly suggested to psychologists to
study mathematical structures. On~page 280 of his \textit{Science and 
Sanity}~\cite{science_and_sanity}, we find a discussion of the importance of the
theory of aggregates and the theory of groups in psychology, something
that will be further examined in Blanco's bi-logic~\cite{blanco,blanco2,rayner,lombardi}.

At variance with the general case~\cite{stealing_general_semantics},
there are some instances in which
the influence of Korzybski's GS on various approaches
is properly acknowledged.  For~example, Ellis acknowledges
Korzybski's influence on his rational emotive behavior
therapy~\cite{ellis}. Almost similarly, Wysong pays the dues of Gestalt
therapy to GS by writing a commentary in
\textit{The Gestalt Journal}~\cite{wysong}. How much Gestalt therapy owes to
GS is also discussed in the thesis of Allen Richard
Barlow~\cite{barlow}, which is downloadable from
The University of Wollongong Thesis Collection online.
One of many counter-examples~\cite{stealing_general_semantics} is given
by family therapy~\cite{minuchin,bowen}, where it is stressed that
one must be aware of abstractions leading to disregarding
the wholeness of processes~\cite{minuchin} (non-elementalism~\cite{science_and_sanity,drive_yourself_sane}),
and the difference between
the verbal and the non-verbal~\cite{bowen} is also underlined, but~without citing GS.
Hence, GS may be considered (either directly or indirectly)
as the hidden root of various therapeutic~practices.

Given the above discussion, it is not difficult to see the logical
connections between GS and QM. If~we consider that
scientific theories are ``maps'' of reality, with~classical theories
providing a first level of abstraction, QM is clearly characterized
by a second level of abstraction. QM does not provide laws for the dynamics
of models of phenomena. QM gives laws for the probability amplitudes
that models of phenomena have a
certain dynamics~\cite{ballentine,weinberg}, i.e.,~QM provides laws
for models of models.
GS classifies this as the self-reflexiveness of the language.
From this perspective, we can consider GS as an application of
certain QM concepts to clinical~psychology.

\subsection{Pauli and Jung's~Synchronicity}
\label{sec:jung}

The goal of the collaboration between Jung and Pauli was
to find a unified view of reality in terms of both the psychological
and physical point of view.
Jung's approach to the psyche was based on certain
{\rm in-forming} (in the sense of having the power of giving ``form'')
structures that he called archetypes~\cite{archetype}.
As universal regulators of the psyche, archetypes transcended the
individual and belonged to a collective unconscious, common to all
humankind.
From the point of view of physics, this can be deemed
much less mysterious than how it sounds. Human ideas are formulated
by brains that share a common physical structure. Although~it has
a great flexibility, such a structure may be expected to constrain
the type of ideas that can be formulated. In~other words, any idea that
can be potentially formulated (which will be called ``archetype'')
must belong to the set of all ideas
permitted by the common brain structure of humans. If~we now
call such a set ``collective unconscious'', we might give a biological
justification to Jung's theory~\cite{percival}.

Pauli was one of the founders of QM. He interpreted QM in terms of
the concept of statistical causality. This facilitated the collaboration
with Jung. He explained to Jung that QM is
about `forms', e.g., wave amplitudes,
and it is also intrinsically probabilistic. While the causality
of the classical world requires the exchange of physical quantities
(such as energy, momentum, angular momentum, and so on), statistical
causality describes a new type of non-local
correlation between systems.
Such a new type of correlation is typically quantum in nature 
and is based on synchronistic events, i.e.,~random
non-local coincidences
~\cite{pauli,fuchs,lindorff,synchronicity,jung-pauli}.
One example is given by the absorption of a quantum of energy from light.
The energy of the quantum is proportional to the frequency $\nu$
of the radiation
and is spread out along the whole wavefront. However, because~of quantum
mechanical fluctuations, the~energy $h\nu$, where $h$ is Planck's constant,
can {instantaneously} disappear
from the wavefront and be instantaneously transferred to a particle
with resonant de Broglie frequency $\nu=p/2mh$, where $p$ is the momentum
of the particle and $m$ is its mass~\cite{ballentine,weinberg}.
The time at which the quantum of energy $h\nu$ disappears from the
wavefront and instantaneously reappears as belonging to the particle's
kinetic energy (synchronistic events) is completely random.
It seems a pure `coincidence'.
In practice, the~ formalism of creation and destruction operators
of quantum field theory~\cite{zee,mandl} describes quantum processes
in terms of random disappearances and reappearances of energy
quanta to and from different systems. However,~everything is probabilistic.
There is no reason for something to happen at a given time
or for the quanta to reappear in the energy of one particle
instead of another one. The~word `coincidence' can indeed
be considered~appropriate.

Analogously, Jung considered random coincidences in the classical world as
the analogue of the statistical causality in the quantum world. Moreover,
in Jung's theory, random coincidences
were also the origin of subjective meaning.
This also means that the organizing principle of reality,
which Jung called synchronicity, is found in meaningful coincidences.
Afterwards, the~concept of synchronicity was further
generalized to include acausal correlations without any psychological component.
We can conclude that Jung's synchronicity
reflects the quantum--classical nature of the~world.

\subsection{The Bi-Logical Structure of~Psychology}
\label{sec:blanco}

Korzybiski's GS~\cite{science_and_sanity} proposes a new psychology founded
on mathematical structures and, to~this end, briefly dealt with both
set and group theory. However, it is only in the work of Blanco~\cite{blanco,blanco2,rayner,lombardi} that these ideas are fully exploited
in order to generalize Freud's formulation of the unconscious.
Whereas Freud defined the unconscious in a qualitative way, i.e.,
what is hidden and repressed in the psyche, Blanco describes it as a bipartite
structure.
Such a bipartite structure has one side that is asymmetric (which we
may call Aristotelian by following GS language), pertaining man's
common-day experience, and~another side that is symmetric (which we
may call non-Aristotelian), where space and time do not exist and
the logical principle of non-contradiction is no longer valid.
Blanco stated that both logics are at work in the human psyche
~\cite{blanco,blanco2,rayner,lombardi} and that clinical practice
must accurately take into account this~point.

Blanco's and GS's conceptual structures share concepts taken from QM.
However, whereas GS is fully non-Aristotelian (without any form of classical-like
logic attached to it), Blanco's bi-logic has an Aristotelian component
(congruent with a classical worldview)
and another non-Aristotelian component (in agreement with the logic of QM).
Taking both aspects into account, we conclude that Blanco's bi-logic
~\cite{blanco,blanco2,rayner,lombardi}
formulates a QC conceptual perspective of the psyche, reflecting the
QC nature of the phenomenological world.
A full acknowledgement of this parallelism and its possible
consequences on clinical practice are a matter of novel research.

\section{Schr\"odinger's `Order from Order' and Jordan's Quantum~Amplification}
\label{sec:jordan}

In this section, we want to put into evidence two quantum mechanical
effects that we consider fundamental for biological matter~\cite{mcfadden-book,mcfadden-alkhalili} in~general
and for the brain in~particular.
Our quantum--classical model can be designed so that it manifestly
incorporates them. One such effect is Schr\"odinger's `order from order'
~\cite{what_is_life},
and the other is Pascual Jordan's quantum amplification
~\cite{beyler,beyler2,mcfadden-alkhalili}.

Classical mechanics applied to biological matter,
including the brain, exploits
statistical fluctuations, i.e.,~the
mechanism that Schr\"odinger called 'order from disorder' \cite{what_is_life}. 
What Schr\"odinger actually wanted to express with the
expression 'order from disorder' is that there are some
ordered macroscopic structures that can arise from the statistical disorder
at the microscopic level. In~truth, `microscopic statistical disorder'
is a misnomer that stands for the great number of microscopic states
that correspond to the same macroscopic state~\cite{callen,blundell}.
Von Neumann entropy (and its quantum--classical generalization defined
in terms of the mixed Weyl of the statistical operator)
is a property of the macrostates given in terms of the probability
of microstates~\cite{ohya,heusler}.
The belief that the passage to macroscopic `order' is associated
with an entropy decrease is mistaken.
The first reason is that macroscopic `order' is somewhat
an anthropomorphic concept that can only be defined once
some macroscopic variables are chosen. On~the contrary,
microscopic order is physical since it is defined in
terms of the number of microstates that are compatible
with the macroscopic constraints. A~system must be
considered microscopically ordered if there is a small
number of states associated to the macroscopic constraints.
In agreement with the third law of thermodynamics, for~example,
this takes place at $T=0$, where there is only one accessible
microstate and the system is maximally ordered on the microscopic
level. Another example is given by the phenomenon of 
re-entrant phase transitions
~\cite{inverse_melting,inverse_melting2,inverse_melting3,inverse_melting4},
where the macroscopic `ordered' phase has a higher entropy than the microscopic
`disordered' one because of the unfreezing of certain DOF.
Irreversible microscopic dynamics, such as diffusive motion,
does not conserve the number of accessible microstates of the system
conditioned by the macroscopic constraints and thus leads to an
increase in \mbox{entropy~\cite{ohya,heusler}}.
This is the essence of Schr\"odinger's 'order from disorder' mechanism
~\cite{what_is_life}:
in our macrocosm, we are surrounded by structures that we classify
as ordered but that are based on microscopic disorder in agreement with the
second law of~thermodynamics.

As discussed by Schr\"odinger, an~'order from disorder' mechanism
can explain neither the stability of biological information nor
the synchronization of molecular processes.
To explain living matter, 
Schr\"odinger proposed a second mechanism
that he named 'order from order'.
The mechanism of 'order from order' is basically founded on
a quantum-mechanical zero-temperature clockwork
in agreement with the third law of thermodynamics~\cite{what_is_life}.
Only solid forms of matter allow for quantum clockworks to exist
in a high-temperature disordered biological environment.
This is caused by the existence of energy gaps protecting, e.g.,
long-wavelength electronic wavefunctions in solids.
This idea led Schr\"odinger to predict that an aperiodic
solid (ultimately identified with DNA~\cite{dna,dna2})
would contain, in a stable manner, the information needed
by the living organism to survive entropic decay.
Currently, the~idea has become more general
and is not limited to solid structures as shields
from molecular disorder. One example is found in the Orch OR theory,
according to which quantum effects are protected inside hydrophobic regions
of biological microstructures
~\cite{microtubulines_channels,microtubulines_channels-2}.
Another mechanism used to protect the quantum clockwork is provided
by rigid boundaries enclosing quantum variables
~\cite{mukamel,mukamel2,pyp}.

Pascual Jordan's idea~\cite{beyler,beyler2,mcfadden-alkhalili}
about how quantum mechanics can steer the dynamics of a classical biological
environment also works for systems with no genetic code
and, as~such, is more general than the `order from order' mechanism.
Jordan introduced the concept of quantum amplification. Basically,
this consists of interpreting the stochastic collapses
~\cite{stapp,vonneumann} of a quantum state
as a way to funnel information from a smaller-scale level to
a classical higher-scale level, i.e.,~an amplification.
Even a small number of quantum variables can 
produce a quantum amplification
through the collapse of their state, and~can have a significant effect on the dynamics of large classical systems.
Quantum amplification is also at work in
non-adiabatic transitions between energy states of a quantum \mbox{subsystem~\cite{osborn,martens,donoso,as3,as13,as7,ak}}
through the back-reaction onto a classical environment,
which the quantum subsystem interacts~with.

Both the mechanisms of `order from order'
and quantum amplification
are in some way present in the quantum approaches to brain dynamics
~\cite{mcfadden,mcfadden2,mcfadden3,pockett,pockett2,pockett3,
liboff,liboff2,frohlich,ph,ph-2,ph-3,
penrose,penrose2,penrose3,penrose4,penrose5,
microtubulines_channels,microtubulines_channels-2,fisher,fisher2,
fisher4,kerskens,qc-bneuro-comp,vitiello1995,pessa-vitiello,alfinito-vitiello,
freeman-vitiello,
freeman-vitiello2,freeman-vitiello3,vitiello-fractals,vitiello-cortex,
sabbadini-vitiello}
that we are going to discuss in the following.
However, since we are explicitly identifying them, we will be able to
design our quantum--classical model in order to take them into account
in a general~way.

In addition to the mechanisms of `order from order'
~\cite{what_is_life} and quantum amplification~\cite{beyler,beyler2,mcfadden-alkhalili},
currently, quantum informational approaches~\cite{nielsen-chuang,jaeger}
are routinely invoked to understand, e.g.,~condensed matter
~\cite{wen} systems.
However, we believe that the idea of considering a biological systems
akin to a quantum computing machine is somewhat controversial.
As we try to explain in this paper, it is even more controversial
than quantum biological theory based on the properties of
a few quantum particles, on~the 'order from order' mechanism
~\cite{what_is_life}, and~on the
quantum amplification process~\cite{beyler,beyler2,mcfadden-alkhalili}.
The reason for this is that, despite a few suggestions~\cite{no-ent-comp,no-ent-comp2,no-ent-comp3,no-ent-comp4},
entanglement is considered to be a necessary resource for
quantum computers~\cite{nielsen-chuang,jaeger}.
However, the~main resource of entanglement 
is the persistent quantum coherence
that, because~of thermal disorder and decoherence, cannot
commonly last for time intervals long enough to perform robust
quantum computations inside the brain.
Nonetheless, it has been proposed to interpret brain processes
as a form of quantum computing
~\cite{microtubulines_channels,fisher,fisher2,kerskens,qc-bneuro-comp}.
It is still a matter of debate that
such a type of computation can take place in the brain
~\cite{microtubulines_channels} nothwithstanding
decoherence~\cite{joos,zurek-2003,tegmark}.

Hence, we think that quantum processes in the brain
can be studied from the perspective of quantum biology
~\cite{mcfadden,mcfadden-alkhalili} with a
somewhat less controversial approach.
Henceforth, we will not pursue the quantum informational perspective
~\cite{nielsen-chuang,jaeger}.

\section{Electromagnetic Fields in the~Brain}
\label{sec:emf}

In this section, we discuss the EMF field approach to brain
dynamics~\cite{mcfadden,mcfadden2,mcfadden3,pockett,pockett2,pockett3,liboff}
and propose its generalization to include {those}
quantum effects that can be influenced
by the various forms of electromagnetic brain stimulation
~\cite{riddle,riddle2,abubaker,croce,caruana,nitsche-paulus,
stagg-nitsche,papazova,yavari}
and that can be observed by \mbox{EEG~\cite{pockett3,croce,nunez}}.

The role of EMFs in bridging space and time scales is very important
~\cite{mcfadden,mcfadden2,mcfadden3,pockett,pockett2,pockett3,liboff,
liboff2,frohlich}. Brain states are routinely studied via
computer simulation~\cite{deco} and various noninvasive
stimulation techniques, such as 
alternating current stimulation (ACS)
\cite{riddle,riddle2,abubaker,croce,caruana}
and transcranial direct-current stimulation (tDCS)
\cite{nitsche-paulus,stagg-nitsche}.
In particular, tDCS is one of the most investigated methods in the field
of non-invasive brain stimulation. It modulates the excitability of the
cerebral cortex with direct electrical currents 
(1 $\approx$ 2 mA~\cite{papazova})
delivered via two or more electrodes of opposite polarities
(i.e., anode and cathode)
placed on the scalp. tDCS modulates resting neuronal membrane potentials
at sub-threshold levels~\cite{nitsche-paulus}, with~anodal and cathodal
stimulation  increasing and decreasing cortical excitability, respectively~\cite{stagg-nitsche}. Although~their tDCS-induced physiological
mechanisms are not yet fully understood, it is assumed that effects
are based on long-term potentiation (LTP) and long-term depression-like
(LTD) mechanisms~\cite{stagg-nitsche,yavari}.

In the history of brain research, it was assumed
that higher brain functions, such as learning and memory,
arise from electrical impulses passing through neurons.
The physical explanation of permanent information storing 
was assigned to multiple reflections of impulses through
neuronal circuits~\cite{mcculloch,caianiello}. This idea is basically 
exemplified by the Hodgkin--Huxley model~\cite{hodgkin-huxley,hodgkin-huxley2}.
Despite its undoubted success,
some limitations of this model have been discussed
~\cite{hh_useful,all-or-none,bodeng,schmitt,arbib}
and possible generalizations have been suggested~\cite{agnati,bower,strassberg}.
From our perspective, the~discussion regarding the role of ion channels
~\cite{bower,strassberg} is particularly important. 
The relevance of quantum effects for charge transport in ion channels
has been strongly supported in Refs.~\cite{ganim,plenio_ionc}.
Typically, this implies that ion channels' selectivity
may be founded on quantum dynamics~\cite{summhammer,salari_2011,bernroider}.
Ultimately, this line of research impinges on possible extensions of
the Hodgkin--Huxley model, not only to take into account the description
of ion channels' conductance but also to incorporate quantum effects
~\cite{moradi}.

The idea that other physical agents, rather that the sole dynamics of
neural networks, must be invoked to describe 
highly coordinated brain activity is not new
~\cite{frohlich,koch,martinez-banaclocha,pinotsis}.
Electric charges (e.g.,~electrons, protons, ions),
together with their associated currents,
are the sources of EMFs~\cite{mcfadden,mcfadden2,mcfadden3,
pockett,pockett2,pockett3,liboff,liboff2}.
In turn, these EMFs interact with water dipoles
and also influence van der Waals and Casimir interactions among 
brain macromolecules. ACS has shown the importance of
EMFs in the brain~\cite{riddle,riddle2,abubaker,croce,caruana}, and
tDCS of human subjects~\cite{nitsche-paulus,stagg-nitsche}
has shown the importance of
both cognition processes and psychological state changes, which 
can be modulated.
For instance, anodal (excitatory) tDCS of the  prefrontal cortex boosts
affective memory, such as fear extinction learning
~\cite{vicario2020a,ney2021,markovic2021}.
Moreover, the~cathodal (i.e., inhibitory) stimulation of the tongue motor
neurons of the primary motor cortex reduces appetite
~\cite{vicario2020b}.

The working of tDCS might be understood through a mechanical analogy.
The complex dynamics of brain EMFs can be reduced to the time evolution
of their sources. Such dynamics can be mapped
onto that of a harmonic spring mattress.
Within this pictorial description, tDCS can be equated
to the nonlinear effect generated by the application of
a constant pressure to
specific extended regions of the spring mattress. The~
applied pressure changes the harmonic dynamics of the mattress
so that oscillations with principal frequencies (phonons) 
scatter with each other.
This mechanical model might be useful for performing computer simulations of
certain processes that are observed in tDCS. We note that the same model
has been used to give a pictorial representation of 
quantum fields~\cite{zee}.
Both ACS and tDCS provide evidence that brain EMFs are not ephemeral;
they are correlated to the dynamics of their sources, but~also
react back and influence both cognitive functions and~emotions.

When studying brain dynamics on the mesoscopic scale of EMFs, it may seem
that there is no necessity to invoke any quantum effect.
The original EMF approach was formulated only in terms of classical physics
~\cite{mcfadden,mcfadden2,mcfadden3,pockett,pockett2,
pockett3,liboff, liboff2,frohlich}.
Nevertheless, our analysis below can elucidate the fundamental
quantum coherent
properties of the microscopic EM fields invoked by such an approach.
~\cite{mukamel,mukamel2,pyp,zee,mandl,romijn,rozyk-myrta}.
Observable coherent EMFs have, by definition, a well-defined phase.
Quantum mechanically, phase $\Phi$ and
photon number $N$ are conjugate variables.
This implies that they obey the indeterminacy relation
\be
\Delta\Phi\Delta N \ge 1
\;. \label{eq:phase}
\ee 
{According} 
 to Equation~(\ref{eq:phase}), when the number of quanta of the 
photon field $N$ is not fixed and $\Delta N$ can be large,
it follows that the phase $\Phi$ is well determined and
the quantum photon field is~coherent.

The only way for the number of photons $N$ to fluctuate
is that photons are continuously absorbed and re-emitted.
In other words, coherent EMFs are `composed' of virtual
photons~\cite{romijn}, e.g.,
packets of energy in momentum space whose existence is ephemeral.
Interestingly, experimental evidence shows that dendrimers can act as
a trap for photons~\cite{mukamel,mukamel2}.
According to quantum electrodynamics \cite{mandl},
a trapped photon can be represented in terms of
virtual photons continuously emitted and re-absorbed between
fermions.
This picture can be developed considering that, in~terms
of Feynman diagrams, a~photon line connecting two fermion lines
is a virtual photon describing M\o ller scattering~\cite{mandl}.
Thus, an~exchange of virtual photons along the time direction between the two
fermion lines, generating a so-called ``ladder'' diagram
~\cite{mahan,mattuck},
may very well be considered the microscopic picture of a trapped~photon.

\section{Penrose and Hameroff's Orch~OR}
\label{sec:penrose}

Orch OR theory
~\cite{ph,ph-2,ph-3,penrose,penrose2,penrose3,penrose4,penrose5,
microtubulines_channels,microtubulines_channels-2,fisher,fisher2,
fisher4,kerskens,qc-bneuro-comp} provides a detailed molecular mechanism
for the time evolution of brain states.
According to Orch OR theory~\cite{ph,ph-2,ph-3,penrose,penrose2,
penrose3,penrose4,penrose5,microtubulines_channels,
microtubulines_channels-2,fisher,fisher2,fisher4,
kerskens,qc-bneuro-comp}, quantum effects in tubulin proteins
(which are organized in arrays of microtubules inside the cytoplasm of 
brain cells) play an important role in brain function.
Quantum dynamics of the electronic orbitals of carbon rings
inside tubulins, time evolution of the nuclear spins,
quantum energy transport among microtubules,
and the spontaneous collapse of microtubules' wave function
are the main ingredients of this theory.
Upon collapse of the wave function, classical brain dynamics ensues.
{For example, w}e 
{look at} charge{s'}
and mass{es'} tunneling
{as an event arising from the collapse
of these variables' wavefunctions.}
{In turn, such a collapse induces} the collapse of
the environment{'s} and
EMF{'s quantum states},
triggering chemical reactions, diffusion processes,
{macroscopic currents,} and so~on.

One peculiar characteristic of Orch OR is that neurons are not
considered to be the fundamental units of information processing~\cite{frohlich}.
Instead, in~Orch OR, it is proposed that information processing takes place
in ordered arrays of microtubules inside the cell.
This idea slowly took form during the 1980s and the first part
of the 1990s when
Hameroff noticed the effects of anesthetics on networks of microtubules
inside the cell.
In a series of papers, Hameroff~et~al.
\cite{net,net2,net3,net4,net5,net6}
proposed that some kind of digital computation was taking place
in arrays of microtubules. Such a computation was based on nonlinear
electrodynamic effects~\cite{net,net2,net3,net4,net5,net6}.
However, the~question of how the results of local digital
calculations could be efficiently transferred
between distant brain regions by classical diffusive mechanisms
remained.
Hence, Hameroff started his search for different mechanisms.
On a different path, looking for a fundamental explanation of wave
function collapse in QM,
Penrose elaborated the theory of objective reduction (OR)
\cite{penrose,penrose2,penrose3,penrose4,penrose5}.

In the standard interpretation of QM, the~collapse of
the wave function, {\rm i.e.}, the~transition from the world
of possibilities 
to that of classical events~\cite{kastner,kastner2}, is explained only
through the stochastic interaction of quantum systems with
classical ones.
The collapse of the wave function is called the `measurement' process
because of the interaction with a classical system~\cite{infamous_boundary}.
It is not explained within the theory but it is assumed as a postulate.
OR proposes that the superposition of different
stationary mass distributions becomes unstable because of quantum
gravitational effects, and, beyond~a certain time interval threshold,
it naturally collapses according to the standard probabilistic rules of QM,
but without any external intervention of a ``measuring instrument''.
A simple way to discuss this process is to consider 
\be
\omega_{\rm Bohr}=\Delta E /\hbar\;,
\label{eq:ome_penrose}
\ee
as the Bohr frequency of the energy eigenvalues
of two eigenstates involved in a certain superposition.
Penrose gives a number of reasons for why the superposition must become 
unstable in the presence of quantum gravitational effects.
The lifetime of the superposition is given by
\be
\tau \approx \frac{h}{\Delta E} \;.
\label{eq:tau_penrose}
\ee
{Looking} at Equations~(\ref{eq:ome_penrose}) and (\ref{eq:tau_penrose}),
one might say that, in~a certain sense, the~deterministic
time evolution of the gravitational field acts as
the instrument measuring the superposition.
However, according to Penrose~\cite{penrose,penrose2,penrose3,penrose4,penrose5},
there is an important difference between the measurement
of the superposition by a classical instrument and
by a quantum gravitational field.
A measurement performed by a quantum gravitational field
is still a fully quantum mechanical process
and, as such, is intrinsically random and absolutely
non-computable.
Penrose considered that brain dynamics is interspersed with 
discrete events
(see Ref.~\cite{pockett2} for experimental support of this idea).
On a phenomenological basis, such events
parallel the discontinuity of wakefulness and awareness~\cite{pockett2}
and other rhythmic phenomena in the brain.
Penrose identified discrete events in the brain with a series
of wave function collapses.
Between one collapse and the other, the~brain can evolve
coherently so that new superpositions are formed.
We note that such a coherent evolution of the wave function,
interrupted by quantum gravitational collapses, is reminiscent of both
piecewise deterministic processes in open quantum systems~\cite{ops}
and nonadiabatic dynamics of the QC system on an 
adiabatic basis
~\cite{as3,as13,as7,ak}.

While Penrose put forth the idea that OR could have an important role
in brain dynamics,
Hameroff fleshed out the detailed biomolecular mechanisms.
Inside each tubulin protein making up a given microtubule, Hameroff
hypothesized the existence of quantum matter systems able to support
stable quantum dynamics in between OR events.
One example is given by carbon rings and their delocalized molecular
orbitals, which can evolve coherently in a superposition of states.
The carbon rings are pushed by hydrophobic forces into the
tubulin's interior,
shielding them from the decoherence~\cite{joos,zurek-2003}
caused by the polar environment outside the protein.
The carbon rings form helical structures inside each microtubule.
They also create oriented arrangements that can act as quantum 
channels~\cite{microtubulines_channels,microtubulines_channels-2}
through which quantum signals travel
among the lattice of microtubules inside the cell's~cytoskeleton.

Various types of quantum oscillators are therefore found
in microtubules' ordered structures, e.g.,
time-dependent electric fields arising from the dynamic polarization of
molecular charges (which produce van der Waal and Casimir--Polder forces),
magnetic fields originating from electron spin dynamics, etc.
Notably, it has also been suggested~\cite{fisher,fisher2,fisher4}
that nuclear spins can play an important role in Orch OR theory
since they are shielded from decoherence for longer time intervals than
other quantum systems in the brain.
Recently, this theory~\cite{fisher,fisher2,fisher4}
has gained experimental support~\cite{kerskens}.
The frequencies of all such quantum oscillators range from kilohertz to
terahertz. 
Orch OR theory requires the feedback~\cite{qc-bneuro-comp} between
the quantum coherent evolution of microtubules
and, for~example, the~classical dynamics
of microtubule-associated proteins (MAPs) \cite{maps,maps-2}.
Such classical dynamics concerns the classical
evolution of MAPs~\cite{maps,maps-2} and CAMKII
~\cite{camkII,camkII-2,camkII-3,camkII-4,camkII-5,camkII-6}, {viz.}
the direction of motion, the~place where MAPs and CAMKII halt their motion,
the case in which they interact or do not interact
with the tubulins, and~the precise time when they interact.
According to the Orch OR theory
~\cite{ph,ph-2,ph-3,penrose,penrose2,
penrose3,penrose4,penrose5,microtubulines_channels,
microtubulines_channels-2,fisher,fisher2,fisher4,kerskens,
qc-bneuro-comp}, the~coherent evolution of the microtubule's wave function
and its OR determine all detailed molecular events.
However, we must note that the possibility that extended brain regions
may be free from 
decoherence~\cite{joos,zurek-2003,tegmark,microtubulines_channels} is rather~controversial.

Lately, there has been a convergence of ideas between the approach
to brain dynamics via quantum EMFs~\cite{mukamel2,mukamel,romijn}
and Orch OR~\cite{neurophotonics}. The~physical process underlying
quantum signaling in Orch OR has been assumed to be photon emission.
Due to the work of Alexander Gurwitsch,
it has been known since the beginning of the 20th century
that tissues inside the body emit biophotons~\cite{popp,popp-2,cifra}.
Such biophotons may be supported by the hydrophobic interior region
of tubulins, where tryptophanes, with~their indole rings of $\pi$ electron
orbitals forming optically active molecular orbitals, are found. 
The packing of indole rings may give rise
to resonant energy transfer between molecular orbitals~\cite{neurophotonics}
much in the same way that F\"orster resonant energy transfer takes place 
between close chromophores. Kurian et al. \cite{kurian}
represented the microtubule as a chain of two-level systems
and calculated the coupling constants in the Hamiltonian
by means of molecular dynamics simulations 
and quantum chemical calculations.
Exciton propagation was performed by means of the Haken and Strobl
method~\cite{haken}.
Their main result is that
energy transfer occurs on a length scale of at least microns.
What is even more interesting from the quantum optical perspective is that
Kurian et al.'s simulation~\cite{kurian} does not consider
the geometric structure of the left-handed helixes of microtubule 
in mammals.
There are reasons to believe that super-radiance can be important
in such complicated geometric arrangements~\cite{superrad,superrad2,superrad3}.
Very recently, the~experimental study of Kalra et al. \cite{kalra}
found that photonic energy transfer in microtubules occurs over 6.6 nm,
cannot be explained in terms of F\"oster theory, and is damped by anesthetics.
The idea that electromagnetic resonance is the fundamental mechanism of 
communications among molecules was first proposed by Veljkovic~et~al.,
who also suggested that such a mechanism could provide
long-range effective communication~\cite{veljkovic}.
At this stage, we believe that a unification of the EMF
and Orch OR theories of brain dynamics
is conceptually very probable~\cite{rozyk-myrta}.

Nevertheless, the Orch OR model remains very controversial.
It is based on quantum gravitational effects used to objectively
induce the wave function collapse by using only
a provisional theory of quantum gravity~\cite{ph,ph-2,ph-3}.

\section{The Dissipative Quantum Model of~Brain}
\label{sec:umezawa}

The precursor of DQMB~\cite{vitiello1995} was
the seminal paper~\cite{umezawa} of Ricciardi and Umezawa,
where the quantum field theory model of brain (QFTMB) was introduced
~\cite{thermofd,thermofd2}.
An interacting QFT can naturally describe the creations
of dynamical correlations.
Whenever a quantum field has an average value different from zero
in the vacuum, the~vacuum state will no longer be unique.
Instead, there will be different vacua and each of them
will spontaneously break the symmetry of the Hamiltonian density
~\cite{nambu,goldstone,goldsalw}.
In order to compensate for the SSB, the~proliferation
of bosonic modes, establishing
long-range correlations with the local configurations of the field, 
sets in. The~symmetry-breaking mechanism in the
QFTMB~\cite{umezawa}
can qualitatively describe both long-term memory storage in the ground states
with broken symmetry and long-range correlations between
distant clusters of neurons by means of the Nambu--Golstone bosons.
Nambu--Goldstone bosons also act as the agents for
memory retrieval~\cite{umezawa} while
excited energy states of the field describe short-term~memory.

In DQMB, the dissipation is ascribed to excited thermal states, 
which are represented through doubling
the number of fields
according to thermo field dynamics~\cite{thermofd,thermofd2}.
DQMB also predicts that
long-range correlations between distant
excited areas of the brains
do not occur via chemical transport but by means
of Nambu--Goldstone \mbox{bosons~\cite{nambu,goldstone,goldsalw}}.
One example of such long-distance correlations
is observed when the brain
is locally stimulated. In~this case, there
is experimental evidence
~\cite{riddle,riddle2,abubaker,croce,caruana,nitsche-paulus,stagg-nitsche}
that the response is given by 
simultaneous excitations in several regions~\cite{clustering,clustering2},
which are far from one another.
In DQMB, quantum coherent fields interact with  classical
neurons and glia cells.
DQMB presents us with a hybrid description where
memory storage finds a quantum explanation and biochemical reactions find
a classical one. Such a hybrid description requires to coarse-grain
the classical degrees of freedom (DOFs) and to describe them in terms
of some kind of waves. Only at this level of description is
 it possible to formulate the interaction between the Nambu--Goldstone
bosons~\cite{nambu,goldstone,goldsalw}, the condensed quantum
field predicted by the model, and the classical waves,
much in the same way that phonons in an ordered solid interact with acoustic waves
~\cite{vitiello1995,pessa-vitiello,alfinito-vitiello, freeman-vitiello,
freeman-vitiello2,freeman-vitiello3,vitiello-fractals,
vitiello-cortex,sabbadini-vitiello}.

Due to its mesoscopic nature, DQMB does not aim at describing 
the behavior of the molecular constituents of the brain
with atomistic detail,
e.g.,~neurons, glia cells, membranes, neurotransmitters, or other
macromolecules. Today, we know that all of these structures
form brain clusters~\cite{clustering,clustering2}
that, once stimulated
~\cite{riddle,riddle2,abubaker,croce,caruana,nitsche-paulus,stagg-nitsche},
can influence human behavior~\cite{croce,caruana}.
Since normal mesoscopic brain dynamics is not chaotic,
the brain's response to stimuli cannot be expected to depend
on the number $N$ of the fundamental constituents of the clusters.
If $N$ is not fixed, Equation~(\ref{eq:phase}) is valid,
the phase $\Phi$ of the matter field is well defined,
and the matter field will be coherent.
{Moreover}, DQMB does not specify the physical nature 
of the bosonic fields of the brain.  The~bosonic fields in Fourier space
may be identified with the modes of the quantum oscillators
considered in Orch OR theory
~\cite{microtubulines_channels,microtubulines_channels-2}
and discussed in Section~\ref{sec:penrose}.
However, another proposal suggested to interpret the bosonic fields
in terms of the dipoles of water molecules
~\cite{delgiudice,delgiudice+1,preparata,jibu,jibu2}.
According to the theory in Refs.
~\cite{delgiudice,delgiudice+1,preparata}, when water
molecules have a high density, the~approximation of weak coupling
to the electromagnetic vacuum field~\cite{qed-coherence} may not hold.
It has been suggested that water in the cytoplasm is found in
a structured state~\cite{ling}, so the considerations of Refs.
~\cite{delgiudice,delgiudice+1,preparata}
are definitely relevant for brain dynamics.
Since a water molecule is dipolar, a~coherent superposition
of the dipoles of many water molecules can be described by
a coherent quantum dipolar field. Hence, in~this model, it is
the condensation of the quantum dipolar field that produces
a ground state with broken symmetry, i.e.,
many unitarily inequivalent subspaces~\cite{blasone}.
Consequently, Nambu--Goldstone modes arise for restoring symmetry 
at long~range.

In the following, we use 
the Hamiltonian of the noninteracting dipolar wave quanta
of Ref.~\cite{sabbadini-vitiello}
in order to elucidate the theoretical description of
dissipation by means of doubling the DOFs
as described by Umezawa's thermo field dynamics~\cite{thermofd,thermofd2}.
The dynamical variables of DQMB are doubled upon introducing
creation and annihilation operators of physical dipolar wave quanta,
$\hat a^\dag , \hat a$, respectively, and~dual creation and annihilation 
operators of fictitious dipolar wave quanta, 
$\hat v^\dag , \hat v$, respectively. 
For example, the~Hamiltonian of the noninteracting dipolar wave quanta
might be defined as~\cite{sabbadini-vitiello}:
\be
\hat H_0 = \sum_k\hbar\omega_k \left(\hat a_k^\dag\hat a_k 
- \hat v^\dag \hat v_k\right) \; ,
\ee
where $\omega_k$ is the oscillation frequency of each mode.
The interaction between the physical modes and their doubles
can be taken as
\be
\hat H_{\rm I} = i\sum_k\hbar\gamma_k
\left(\hat a_k^\dag\hat v^\dag - \hat a_k\hat v_k\right) \; ,
\ee
where $\gamma_k$ is the damping constant of each mode.
Finally, the~total many-body Hamiltonian of the thermal system is
\be
\hat H = \hat H_0 + \hat H_{\rm I} \label{eq:H-thermo}.
\ee 
A thorough study of the Hamiltonian in Equation~(\ref{eq:H-thermo})
and its associated equations of motion
has led to finding a number of interesting results over
the years~\cite{freeman-vitiello}.

DQMB has been applied by Vitiello and collaborators to study various brain
processes~\cite{alfinito-vitiello}.
Some applications include nonlinear dynamics
~\cite{freeman-vitiello2}, cortical patterns in perception
~\cite{freeman-vitiello3}, the~relation between fractal properties
and coherent states in the brain~\cite{vitiello-fractals},
rhythmic generators in the cortex~\cite{vitiello-cortex},
and correlations of brain regions that are realized through
entanglement~\cite{sabbadini-vitiello}.
DQMB dynamics has also been adopted by
Nishiyama~et~al. in a number of works~\cite{nishiyama,nishiyama2,nishiyama3,nishiyama4}.
As reported in Ref.~\cite{nishiyama}, one notes that
the phenomenon of super-radiance, which is expected to occur in complicated
geometric arrangements of microtubules, also occurs in~DQMB.

\section{The Quantum--Classical Model of~Brain}
\label{sec:qcdmb}

Our aim is to model multi-scale brain dynamics,
explicitly treating classical DOFs and
quantum variables on the same footing.
To this end, our approach considered mixed Weyl symbols of dynamical
variables (represented by operators in the standard formulation of
quantum mechanics) and a mixed Weyl symbol of the statistical operator
(corresponding to the density matrix of the systems in the standard
representation of QM)
\cite{silin,rukhazade,balescu,zhang-balescu,balescu-zhang,aleksandrov,
gerasimenko,boucher,petrina,prezhdo,kapracicco,nielsen,as4,as6,as14,
osborn,martens,donoso,as3,as13,as7,ak}.
In this regard, it is worth remarking that 
we introduced a quantum--classical model of brain dynamics
because, while thermal disorder, decoherence, and~a short de Broglie
{wavelength} require treating the DOFs of massive atoms and molecules
by means of classical mechanics, there are also
intrinsically quantum variables at all temperatures
influencing the dynamics of the classical DOF. Such quantum variables are,
e.g.,~tunneling electrons and protons, nuclear and electron spins,
electronic-state-associated 
phenyl rings, photons, and so on.
Our working hypothesis is that certain brain processes can most naturally
be understood in terms of the above picture.
In the following, we sketch the quantum--classical theory that
we used and introduce a general Hamiltonian model that
can be adapted to perform calculations of specific
phenomena at the interface between the quantum and the classical world.
For example, we plan to study if quantum dynamics can explain
ion channels' selectivity inside the classical membrane of a brain cells.
Quantum particles that are recognized can be characterized in terms of
their vibrational spectra.
Response functions can be calculated depending of the oscillators' frequencies
and the wave vectors, and~these can be compared to
the results of electromagnetic brain stimulation and EEG.
Another example is given by the trapping of photons by neurons' dendrites.
Once trapped, photons should acquire an effective mass and the dynamics of the 
dendrites should be slowed down.
One could expect that some observable effect of or alteration
in neuronal dynamics should follow.
The question is if this investigation could explain
the effects of electromagnetic brain stimulation.
These are just a few examples. In~general,
numerical simulations can be used as thought experiments
for `discovering' the qualitative agreement between a specific model
and real experiments or the causal connection between the
numerical values of the parameters of the model and the behavior of
the model itself.
Comparisons between different Hamiltonian models can also lead
to a better understanding of the essential features of a given phenomenon.
These are the goals of our quantum--classical model of brain dynamics.
Section~\ref{sec:psyche} has logically sustained the choice of adopting
a quantum--classical approach and has pointed us toward the very ambitious
goal of developing models that cannot only go from the microscopic level
to the results of electromagnetic brain stimulation
but that can also reach  the psychological states of the brain.
We intend this as specified by a few macroscopic variables
that can be related to human behavior. If~human behavior can be affected
by electromagnetic brain stimulation and if we are successful in building
models going from the quantum--classical microscopic level to
brain stimulation, then our goal will be reached, even if it is as an overall result
of a research activity whose first results are just those proposed 
in this~article.

We imagine that the brain is described by quantum operators
$(\hat r,\hat p, \hat R, \hat P)$, where $(\hat r,\hat R)$ are position
operators and $(\hat p,\hat P)$ are the respective conjugated
momenta operators.
Now, $(\hat r,\hat p)=\hat x$ corresponds to the brain 
variables with
a long de Broglie wavelength that, for~this reason, must be
treated quantum mechanically, 
whereas $(\hat R, \hat P)=\hat X$ can be treated semi-classically because
of their much shorter de Broglie wavelength.
A partial Wigner transform over
the $(\hat X)$ operators~\cite{kapracicco}
introduces the mixed Weyl symbols $\tilde{\cal O}(X)$ and
$\tilde{\cal W}(X)$ arising from $\hat{\cal O}(\hat x, \hat X)$
and $\rho(\hat x, \hat X )$, respectively.
Please note that the following notation is adopted:
when a quantum operator depends both on quantum 
variables and classical DOFs,
a $\tilde{\phantom{a}}$ is written on it,
whereas, if the quantum operator does not depend on $X$, 
a $\hat{\phantom{a}}$ is used.
No hat is used in the case of a dynamical variable
depending only on $X$.
A practical example of a possible application of this mixed QC representation
can be given when considering molecular orbitals, electron and nuclear spins,
light ions, neurons, glia cells, and~electromagnetic
interactions.
Conformational dynamics of cells may be represented
through phonons, i.e.,~harmonic DOFs.
Other harmonic DOFs can be used to describe coherent EMFs.
The inclusion of non-Harmonic perturbation terms
provides a description of non-trivial interactions among all of the DOFs
of the model.
Zero-point effects on the motion of classical-like DOFs can be described
by means of advanced algorithms that will be explained in the following.
As in the case of DQMB, the~goal is to set up a mesoscale approach
to brain dynamics, noting, however, that, in our case, the QC dynamical variables
are explicitly~represented.

If we now introduce the coordinates of the EMF modes $(Q,\Pi)=\Upsilon$,
a possible model mixed Weyl symbol of the Hamiltonian
$\tilde{\cal H}(X,\Upsilon)$
can be written as
\ba
\tilde{\cal H}(X,\Upsilon)
&=&
\hat{\cal H}_{\rm S} + {\cal H}_{\rm B}(X) + {\cal H}_{\rm F}(\Upsilon)
+ \tilde{\cal V}_{\rm SB}(R)
+ \tilde{\cal V}_{\rm SEM}(Q) 
\label{eq:brain-model}
\ea 
In Equation~(\ref{eq:brain-model}), $\hat{\cal H}_{\rm S}(t)$
is the Hamiltonian operator
of the quantum subsystem with quantum variables $\hat x$.
The phononic Hamiltonian is
\be
{\cal H}_{\rm B}(X)=\sum_{J=1}^{N_{\rm PH}}\left(\frac{P_J^2}{2} +
\frac{(\omega_J^{\rm PH})^2}{2}R_J^2\right) \;,
\ee
where $\omega_J^{\rm PH}$, $J=1,...,N_{\rm PH}$ is the frequency of
each phonon.
Similarly, the EMF Hamiltonian is
\be
{\cal H}_{\rm F}(\Upsilon)=\sum_{K=1}^{N_{\rm EM}}\left(\frac{\Pi_K^2}{2} +
\frac{(\omega_K^{\rm EM})^2}{2}Q_K^2\right) \;,
\ee
where $\omega_K^{\rm EM}$, $K=1,...,N_{\rm EM}$ is the frequency of
the EMF mode.
The interaction operators $\tilde{\cal V}_{\rm SB}(R)$
and $\tilde{\cal V}_{\rm SEM}(Q)$
describe the coupling of the phonons and of the EMF
to the quantum subsystem, respectively. 
Assuming, for simplicity, a bilinear approximation, these can be written as
\ba
\tilde{\cal V}_{\rm SB}(R) &=& -\sum_{J=1}^{N_{\rm PH}} C_J R_J \hat \chi
\\
\tilde{\cal V}_{\rm SEM}(Q) &=& - \sum_{K=1}^{N_{\rm M}} F_K Q_K \hat \zeta \;,
\ea
where the $C_J$ and $F_K$ are the coupling constants of the quantum
operators $\hat \chi$ and $\hat \zeta$, respectively.
The operators $\hat \chi$ and $\hat \zeta$ act on the same space of
$\hat x$.

The dynamics of the mixed Weyl symbol 
$\tilde{\cal O}(X,\Upsilon,t)$ 
of an arbitrary operator $\hat{\cal O}$ is given by a
QC bracket
~\cite{silin,rukhazade,balescu,zhang-balescu,balescu-zhang,aleksandrov,
gerasimenko,boucher,petrina,prezhdo,kapracicco,nielsen,as4,as6,as14,
osborn,martens,donoso,as3,as13,as7,ak}.
The QC bracket is a quasi-Lie bracket
~\cite{nielsen,as4,as6,as14} that breaks the time-translation invariance
of Lie algebras because it does not satisfy the Jacobi relation.
In the case of a system with both phononic and EMF modes,
it can be written by introducing two antisymmetric matrices,
$\mbox{\boldmath$\Omega$}=-\mbox{\boldmath$\Omega$}^{-1}$
and $\mbox{\boldmath$\Lambda$}=-\mbox{\boldmath$\Lambda$}^{-1}$:
\be
\mbox{\boldmath$\Omega$}
=
\left[ \begin{array}{cc} 0 & 1 \\
- 1 & 0 \end{array} \right] 
\label{eq:Omega}
\ee
and 
\be
\mbox{\boldmath$\Lambda$}
=
\left[ \begin{array}{cccc}
{\bf 0} & {\bf 0} & {\bf 1} & {\bf 0} \\ 
{\bf 0} & {\bf 0}  & {\bf 0} & {\bf 1} \\
-{\bf 1} & {\bf 0} & {\bf 0} & {\bf 0} \\
{\bf 0} & -{\bf 1} & {\bf 0} & {\bf 0}
 \end{array} \right] \;.
\label{eq:Lambda}
\ee 
{The} QC equation of motion in the Heisenberg picture reads
\ba
\partial_t\tilde{\cal O}(t)
&=&\frac{i}\hbar
\left[ \begin{array}{cc}
\tilde{\cal H} & \tilde{\cal O}(t)
\end{array} \right]
\mbox{\boldmath$\Omega$}
\left[ \begin{array}{c}
\tilde{\cal H} \\ \tilde{\cal O}(t)
\end{array} \right]
-\frac{1}{2}
\tilde{\cal H}
\overleftarrow{\mbox{\boldmath$\nabla$}^{X,\Upsilon}}
\mbox{\boldmath$\Lambda$}
\overrightarrow{\mbox{\boldmath$\nabla$}^{X,\Upsilon}}
\tilde{\cal O}(t)
\nonumber\\
&+&
\frac{1}{2}
\tilde{\cal O}(t)
\overleftarrow{\mbox{\boldmath$\nabla$}^{X,\Upsilon}}
\mbox{\boldmath$\Lambda$}
\overrightarrow{\mbox{\boldmath$\nabla$}^{X,\Upsilon}}
\tilde{\cal H} \;,
\label{eq:qc-bracket-nve}
\ea
where $\mbox{\boldmath$\nabla$}^{X,\Upsilon}=((\partial/\partial R),
(\partial/\partial Q),(\partial/\partial P),(\partial/\partial \Pi))$
is the phase space gradient~operator.

The lhs of Equation~(\ref{eq:qc-bracket-nve}) defines the quantum--classical
bracket of $\tilde{\cal O}(t)$ with $\tilde{\cal H}$.
The first term in the lhs of Equation~(\ref{eq:qc-bracket-nve}) is
the quantum commutator, whereas the other two terms are Poisson brackets.
All terms are written in matrix form~\cite{as4,as6,as14}.
The super propagator associated to the QC bracket is
\ba
\tilde{\tilde{{\cal U}}} (t)
&=&
\exp\left\{(it/\hbar)
\left[ \begin{array}{cc}
\tilde{\cal H} & \ldots
\end{array} \right]
\mbox{\boldmath$\Omega$}
\left[ \begin{array}{c}
\tilde{\cal H} \\ \ldots
\end{array} \right]
-(t/2)
\left(
\tilde{\cal H}\overleftarrow{\mbox{\boldmath$\nabla$}}^{X,\Upsilon}
\mbox{\boldmath$\Lambda$}
\overrightarrow{\mbox{\boldmath$\nabla$}}^{X,\Upsilon} \ldots\right)
\right.
\nonumber\\
&+&
\left.
(t/2)\left(
\ldots \overleftarrow{\mbox{\boldmath$\nabla$}}^{X,\Upsilon}
\mbox{\boldmath$\Lambda$}
\overrightarrow{\mbox{\boldmath$\nabla$}}^{X,\Upsilon} \tilde{\cal H}
\right)
\right\}
\label{eq:U}
\ea 
{The} super-operator $\tilde{\tilde{{\cal U}}} (t)$ defines the dynamics
of mixed Weyl symbols of standard operators as
\be
\tilde{\cal O}(t)=\tilde{\tilde{{\cal U}}} (t)\tilde{\cal O}
\;, \label{qc-dyna}
\ee
where $\tilde{\cal O}=\tilde{\cal O}(t=0)$.
QC averages are calculated using the formula
\be
\langle \tilde{\cal O}(t)\rangle
={\rm Tr}^\prime\int dXd\Upsilon\; \tilde{\cal W}(X,\Upsilon;t)
\tilde{\cal O}(X,\Upsilon,t) \;.
\label{eq:qc-ave}
\ee 
{In} Equation~(\ref{eq:qc-ave}), the parametric time dependence
of the mixed Weyl symbol of the statistical operator of the system,
$\tilde{\cal W}(X,\Upsilon;t)$, describes possible non-equilibrium
initial conditions. The~formalism presented here can be easily adapted
to more general non-equilibrium situations arising from an explicit
time dependence of the mixed Weyl symbol of the Hamiltonian in
Equation~(\ref{eq:brain-model}). In~such a case, it would be more convenient
to adopt the Schr\"odinger scheme of motion and propagate the mixed Weyl
symbol of the statistical operator. One would also have to take into
account the time ordering of the propagator, something 
that can be implemented by the algorithm~\cite{as13}.
Non-equilibrium dynamics is important if one considers the free energy principle
proposed by Karl {Friston} \cite{friston,sanchez-canizares}.
Recently, such a direction of research has witnessed interesting
developments~\cite{tanaka}.
As for QC correlation functions,
they are defined in the following way:
\be
\langle \tilde{\cal O}_1(t)\tilde{\cal O}_2\rangle
=
{\rm Tr}^\prime\int dXd\Upsilon\; \tilde{\cal W}(X,\Upsilon;t)
\tilde{\cal O}_1(X,\Upsilon,t) \tilde{\cal O}_2(X,\Upsilon) \;.
\label{eq:qc-corr}
\ee 
{The} operator ${\rm Tr}^\prime$ found in Equations~(\ref{eq:qc-ave})
and (\ref{eq:qc-corr}) takes the trace over the quantum operators
$\hat x$, while $\tilde{\cal O}_1$ and $\tilde{\cal O}_2$
are two arbitrary mixed Weyl~symbols.

\subsection*{Constant Temperature Quantum--Classical~Dynamics}

In order to illustrate the advanced techniques for controlling
the temperature of the harmonic modes, we consider 
a simple system with just two phononic modes, with coordinates
$(X_1,X_2)$, and two NHC chains of length one (which is usually
enough to generate ergodic dynamics for stiff harmonic degrees
of freedom) \cite{nhc,b1,b2}. Thus, the~extended phase space point
can be written as
$X^{\rm e}$= $(R_1$,$\eta_1^{(1)},$ $\eta_2^{(1)},$ $R_2,$
$\eta_2^{(1)},$ $\eta_2^{(2)},$ $P_1,$ $P_{\eta_1}^{(1)},$
$P_{\eta_1}^{(2)},$ $P_2,$ $P_{\eta_2}^{(1)},$ $P_{\eta_2}^{(2)});$
consequently, the~extended phase space gradient is
$\mbox{\boldmath$\nabla$}^{\rm e}$=
$((\partial/\partial R_1),$
$(\partial/\partial \eta_1^{(1)}),$
$(\partial/\partial \eta_2^{(1)}),$
$(\partial/\partial R_2),$
$(\partial/\partial \eta_1^{(2)}),$
$(\partial/\partial \eta_2^{(2)})$,
$(\partial/\partial P_1),$
$(\partial/\partial P_{\eta_1}^{(1)}),$
$(\partial/\partial P_{\eta_2}^{(1)})$,
$(\partial/\partial P_2),$
$(\partial/\partial P_{\eta_1}^{(2)})$,
$(\partial/\partial P_{\eta_2}^{(2)}).$

If we now define the antisymmetric matrix $\mbox{\boldmath${\cal R}$}=
-\mbox{\boldmath${\cal R}$}^{-1}$ as
\ba
\mbox{\boldmath${\cal R}$}=
\left[
\begin{array}{cccccccccccc}
0 & 0 & 0 & 0 & 0 & 0 & 1 & 0   & 0 & 0 & 0 & 0 \\
0 & 0 & 0 & 0 & 0 & 0 & 0 & 1   & 0 & 0 & 0 & 0 \\
0 & 0 & 0 & 0 & 0 & 0 & 0 & 0   & 1 & 0 & 0 & 0 \\
0 & 0 & 0 & 0 & 0 & 0 & 0 & 0   & 0 & 1 & 0 & 0 \\
0 & 0 & 0 & 0 & 0 & 0 & 0 & 0   & 0 & 0 & 1 & 0 \\
0 & 0 & 0 & 0 & 0 & 0 & 0 & 0   & 0 & 0 & 0 & 1 \\
-1& 0 & 0 & 0 & 0 & 0 & 0 &-P_1 & 0 & 0 & 0 & 0 \\
0 &-1 & 0 & 0 & 0 & 0 &P_1&0&-P_{M_{\eta_2}}^{(1)}&0&0 & 0 \\
0 & 0 &-1 & 0 & 0 & 0 & 0 & P_{M_{\eta_2}}^{(1)}  & 0 & 0 & 0 & 0 \\
0 & 0 & 0 &-1 & 0 & 0 & 0 & 0  & 0 & 0 &-P_2 & 0 \\
0 & 0 & 0 & 0 &-1 & 0 & 0 & 0  & 0 &P_2& 0 &-P_{\eta_1}^{(2)} \\
0 & 0 & 0 & 0 & 0 &-1 & 0 & 0 & 0 & 0 &P_{\eta_1}^{(2)}&  0 
\;.
\end{array}
\right]
\ea
together with the mixed Weyl symbol of the extended Hamiltonian
\ba
\tilde{\cal H}^{\rm e}(X^{\rm e})
&=&
\hat{\cal H}_{\rm S} + {\cal H}_{\rm B}(X) + \tilde{\cal V}_{\rm SB}(R)
+\sum_{I=1}^2\sum_{L=1}^2\frac{P_{\eta_L}^{(I)}}{2M_{\eta_L}}
+\sum_{I=1}^2\sum_{L=1}^2k_{\rm B}T^{(I)}\eta_L^{(I)}
\label{eq:He}
\;.
\ea
the QC equation of motion at constant temperature can be written
in compact form~\cite{as4,as6,as14} as
\ba
\partial_t\tilde{\cal O}^{\rm e}(t)
&=&\frac{i}\hbar
\left[ \begin{array}{cc}
\tilde{\cal H}^{\rm e} & \tilde{\cal O}^{\rm e}(t)
\end{array} \right]
\mbox{\boldmath$\Omega$}
\left[ \begin{array}{c}
\tilde{\cal H}^{\rm e} \\ \tilde{\cal O}^{\rm e}(t)
\end{array} \right]
-\frac{1}{2}
\tilde{\cal H}^{\rm e}
\overleftarrow{\mbox{\boldmath$\nabla$}^{\rm e}}
\mbox{\boldmath$\cal R$}
\overrightarrow{\mbox{\boldmath$\nabla$}^{\rm e}}
\tilde{\cal O}^{\rm e}(t)
\nonumber\\
&+&
\frac{1}{2}
\tilde{\cal O}^{\rm e}(t)
\overleftarrow{\mbox{\boldmath$\nabla$}^{\rm e}}
\mbox{\boldmath$\cal R$}
\overrightarrow{\mbox{\boldmath$\nabla$}^{\rm e}}
\tilde{\cal H}^{\rm e} \;,
\label{eq:qc-bracket}
\ea
where $\tilde{\cal O}^{\rm e}(t)=\tilde{\cal O}^{\rm e}(X^{\rm e},t)$;
the NHC variables are 
$(\eta_L^{(I)},P_{\eta_L}^{(I)})$, with~$I$ and $L$ running over the
phonons and the coordinates of the chain, respectively; $k_{\rm B}$
is the Boltzmann constant; $T^{(I)}$ is the temperature of each mode;
and $M_{\eta_L^{(I)}}$ are the inertial parameters of the NHC~variables.

Constant temperature averages and correlation functions can be calculated by
choosing the mixed Weyl symbol $\tilde{\cal W}^{\rm e}(X^{\rm e})$
of the statistical operator in extended space as
\ba
\tilde{\cal W}^{\rm e}(X^{\rm e})
&=&
\hat w_{\rm S}
{\cal W}^{\beta}(X)
\prod_{I=1}^2\prod_{L=1}^2
\delta\left(\eta_L^{(I)}\right)\delta\left(P_{\eta_L}^{(I)}\right)
\;
\ea
where $\hat w_{\rm S}$ is the mixed Weyl symbol of the statistical
operator of the quantum subsystem while 
the thermal mixed Weyl symbol of the statistical operators of the 
phonons is
\ba
{\cal W}^{\beta}(X)
&=&\prod_{I=1}^{2}
\frac{\tanh(\beta\omega_I/2)}{2}\exp\left[
-\frac{2\tanh(\beta\omega_I/2)}{\omega_I}
\left(\frac{P_I^2}{2}+\frac{\omega_I^2}{2}R_I^2\right) \right]
\;,
\label{eq:phonon-distribution}
\ea
where $\beta=1/k_{\rm B}T$ and
$\omega_I$ is the frequency of phonon $I$.
If, in the mixed Weyl symbol of the Hamiltonian in Equation~(\ref{eq:He}),
one defines $T^{(I)}=T~\forall I$, then 
the dynamics defined by Equation~(\ref{eq:qc-bracket}) defines
constant-temperature evolution.
Instead, the~choice of \linebreak $T^{(I)} = 1 / k_{\rm B} \beta^{(I)}$ with~
\be 
\beta^{(I)}= \frac{2\tanh(\beta\omega_I/2)}{\omega_I}\;,
\quad \forall I\; .
\label{eq:eff-beta}
\ee
describes a time evolution of the phonons, where zero-point effects are
taken into account.
The structure of the extended QC super-propagator 
$\tilde{\tilde{{\cal U}}}^{\rm e}$ is similar to that displayed
in Equation~(\ref{eq:U}):
\ba
\tilde{\tilde{{\cal U}}}^{\rm e}(t)
&=&
\exp\left\{(it/\hbar)
\left[ \begin{array}{cc}
\tilde{\cal H}^{\rm e} & \ldots
\end{array} \right]
\mbox{\boldmath$\Omega$}
\left[ \begin{array}{c}
\tilde{\cal H}^{\rm e}\\ \ldots
\end{array} \right]
-(t/2)
\left(
\tilde{\cal H}^{\rm e}\overleftarrow{\mbox{\boldmath$\nabla$}}^{\rm e}
\mbox{\boldmath$\cal R$}
\overrightarrow{\mbox{\boldmath$\nabla$}}^{\rm e} \ldots\right)
\right.
\nonumber\\
&+&
\left.
(t/2)\left(
\ldots \overleftarrow{\mbox{\boldmath$\nabla$}}^{\rm e}
\mbox{\boldmath$\cal R$}
\overrightarrow{\mbox{\boldmath$\nabla$}}^{\rm e}\tilde{\cal H}^{\rm e}
\right)
\right\} \;.
\label{eq:Ue}
\ea 
{Since} we are interested in thermal and zero-point QC averages and
correlation functions of non-fictitious dynamical variables,
we must consider mixed Weyl symbols $\tilde{\cal O}(X)$ that, at
$t=0$, do not depend on
the extended phase space point $X^{\rm e}$ but depend on the
non-fictitious phase space point $X$.
However, the~key to temperature control is that the phase space
variable dependence found at $t=0$ is not preserved at $t\neq 0$.
We have $\tilde{\tilde{{\cal U}}}^{\rm e}(t)\tilde{\cal O}(X)
=\tilde{\cal O}(X^{\rm e},t)$.
Finally, we can write the expression for thermal (or zero-point)
QC averages as
\ba
\langle \tilde{\cal O}(X,t)\rangle_{\rm e}
&=&
{\rm Tr}^\prime \int dX^{\rm e}\; {\cal W}^{\rm e}(X^{\rm e})
\tilde{\cal O}(X,t) \;,
\\
\langle \tilde{\cal O}_1(X,t) \tilde{\cal O}_2(X)\rangle_{\rm e}
&=&
{\rm Tr}^\prime \int dX^{\rm e}\; {\cal W}^{\rm e}(X^{\rm e})
\tilde{\cal O}_1(X,t) \tilde{\cal O}_2(X) \;.
\ea


\section{Conclusions}
\label{sec:end}

In this work, we brought to light a
parallel between the 
bipartite structure of human logic and the quantum--classical view 
of physical phenomena.  We discussed that one finds both
Aristotelian logic and non-Aristotelian logic in the human psyche.
Aristotelian logic explains the behavior
of the classical world, whereas non-Aristotelian logic applies to the
quantum world.  We would like to remind the~reader that, in the manuscript,
the word `psyche' means the set of collective brain phenomena
emerging from microscopic cells' dynamics, and is not anything~metaphysical.

We have been motivated by the analogy
with bi-partite logic to propose a quantum--classical model
for studying brain processes. One idea behind this proposal, i.e.,
the need to mix a quantum and classical level of description together,
had already been supported in a somewhat less explicit form by three
theoretical approaches, which we reviewed in the first part of this work.
Two approaches that we reviewed were originally designed
by their authors as theories of consciousness. However, this is not the
perspective from which we looked at them. In~this paper,
we were not interested in describing consciousness. Instead, we 
interpreted these theories in terms of purely physical processes, and it was only
in such a respect that we considered~them.

The very formulation of our model is given in terms of quantum and
classical variables that are treated on the same level. It does not
need to invoke, e.g.,~quantum gravitational effects in the
brain. Instead, the~crux is that the quantum variables 
play the fundamental role of providing a quantum guiding
mechanism for the classical variables that they are coupled with.
Such an idea was originally formulated by Pascual Jordan.
With respect to this, in~order to have quantum effects in brain dynamics,
there is no need to invoke a highly improbable
coherent quantum state of the whole brain at a high temperature.
There are important quantum properties of few-body systems that
are not lost at a high temperature. These were discussed in the text.
What is needed for the quantum biology of the brain was again
suggested long ago by Pascual Jordan: the collapse of the wave function
works as an amplification mechanism acting as a bridge between the quantum
and the classical~world.

We did not perform actual numerical calculation but
we introduced a general quantum--classical Hamiltonian model,
which can be specialized to describe quantum particles and spins
interacting with various types of environments, such as~those found
in neurons and astrocytes, or~at the sub-neuronal level in, e.g.,
tubulins.
Moreover, electromagnetic DOFs can also be described by our model Hamiltonian.
We showed that the quantum--classical theory provides
a statistical mechanic formulation of averages and correlation functions.
In turn, as~is well-known, correlation functions lead to the definition
of response functions. Non-invasive
brain stimulation techniques can provide the numerical data to which
our theory can be~compared.

We took the risk to discuss many
complex ideas using only logic and our scientific knowledge.
We presented a synthesis of subtle concepts, introduced our
quantum--classical model, with~which we plan to take on big
scientific challenges, and declared the direction
that our future work will take. We carried this out with the belief
that science is not only made by numbers, but also made by understanding
and sharing concepts with the community. Subsequently,
such concepts can be discussed and refined,
possibly leading to new advancements.
Our future work will be devoted to interpreting
biochemical processes in the brain in terms of quantum--classical dynamics.
This will also require performing
quantum--classical calculations of neural response functions.
The implications of the interplay between the bipartite structures
of both the world and the psyche will be investigated through
the formulation of quantum--classical models of~decision making.

\end{document}